\documentclass[prl,twocolumn,showpacs,preprintnumbers,amsmath,amssymb]{revtex4}
\usepackage{graphicx}
\usepackage{dcolumn}
\usepackage{bm}

\begin{document}

\title{Overcoming blockade in producing doubly-excited dimers by a single intense pulse and their decay}

\author{\firstname{Ph.~V.} \surname{Demekhin}}
\altaffiliation {philipp.demekhin@pci.uni-heidelberg.de}
\affiliation{Theoretische Chemie, Physikalisch-Chemisches
Institut, Universit\"{a}t  Heidelberg, Im Neuenheimer Feld 229,
D-69120 Heidelberg, Germany}

\author{\firstname{K.} \surname{Gokhberg}}
\affiliation{Theoretische Chemie, Physikalisch-Chemisches Institut, Universit\"{a}t Heidelberg,
Im Neuenheimer Feld 229, D-69120 Heidelberg, Germany}

\author{\firstname{G.} \surname{Jabbari }}
\affiliation{Theoretische Chemie, Physikalisch-Chemisches Institut, Universit\"{a}t Heidelberg,
Im Neuenheimer Feld 229, D-69120 Heidelberg, Germany}

\author{\firstname{S.} \surname{Kopelke}}
\affiliation{Theoretische Chemie, Physikalisch-Chemisches Institut, Universit\"{a}t Heidelberg,
Im Neuenheimer Feld 229, D-69120 Heidelberg, Germany}

\author{\firstname{A.~I. } \surname{Kuleff}}
\affiliation{Theoretische Chemie, Physikalisch-Chemisches Institut, Universit\"{a}t Heidelberg,
Im Neuenheimer Feld 229, D-69120 Heidelberg, Germany}

\author{\firstname{L.~S.} \surname{Cederbaum}}
\affiliation{Theoretische Chemie, Physikalisch-Chemisches Institut, Universit\"{a}t Heidelberg,
Im Neuenheimer Feld 229, D-69120 Heidelberg, Germany}

\begin{abstract}
Excitation of two identical  species in a  cluster by the absorption of two photons of the same energy is strongly suppressed since the excitation of one subunit   blocks the excitation of the other one due to the binding Coulomb interaction. Here, we propose a very efficient way to overcome this blockade in producing doubly-excited  homoatomic clusters by a single intense laser pulse.  For Ne$_2$ it is explicitly demonstrated that the optimal carrier frequency of the pulse is given by  half of the energy of the target state, which allows one to doubly excite more than   half of the dimers at  moderate field intensities. These  dimers then undergo ultrafast interatomic decay bringing one Ne to its ground state and ionizing the other one. The reported \emph{ab initio} electron spectra present reliable predictions for future experiments by  strong laser pulses.
\end{abstract}


\maketitle

Blockade is a general term and encompasses many phenomena in different fields investigated for a long time. Thereby a single particle prevents the flow or excitation of other particles  \cite{Urban_NatPhys09}. The first studies \cite{junct1,junct2} were on single-electron tunneling induced by blockade interactions and date back to the late 1960's   (see also \cite{junct3,junct4,junct5}). Later on, blockade phenomena have been intensively studied in different realizations with atomic Rydberg gases, where the long-range inter-atomic dipole-dipole interactions prevent the excitation of two identical atoms by the absorption of two photons of the same energy \cite{BLREV10}. The excitation blockade was shown to lead to, e.g., spectral line broadening \cite{primer1,primer2,primer3}, enhancement of Penning ionization  \cite{primer4}, and non-Poissonian excitation probability distributions \cite{primer5,primer6}. Recently, blockade in ultracold homoatomic ensembles has attracted considerable attention due to its potential application for quantum information processing \cite{Urban_NatPhys09,BLREV10,Zoller_PRL00,Zoller_PRL01,PhotonBL05,Gaetan_NatPhys09}. It has been shown that the excitation of a single atom in the ensemble can block the excitation of other atoms (even when the atoms are more than 10~$\mu$m apart \cite{Urban_NatPhys09}) due to the dipolar shift of the atomic levels, which brings the driving pulse out of resonance.

The effect of excitation blockade in an ensemble of atoms is particularly pronounced when these atoms form a cluster where the excited states are also shifted by the binding interaction. Particularly suitable systems to study the phenomenon are rare-gas clusters since the bonding is relatively weak and the constituents preserve, to a large extent, their atomic character. At the same time, rare-gas clusters are  easily amenable to experiments: they can be generated in a wide range of different sizes \cite{clust1,clust2},  from a few up to millions of atoms, allowing to  study a transformation of electronic properties  from an atom to the solid. However,  even the first excited states of rare-gas atoms lie in the UV range and cannot be accessed by a single photon with conventional optical lasers utilized in the mentioned above experiments with  cold Rydberg atoms.

The advent of the new generation of light sources, like free electron lasers \cite{FLASH,FERMI} and   high-order harmonic generation setups \cite{highharm1,highharm2}  allow one  to produce ultrashort and intense coherent laser pulses of high frequencies. Exposed to very strong high-frequency pulses, clusters  may absorb a large number of photons creating differently charged ions \cite{Saalmann06} even if the energy of a single photon is not sufficient to directly ionize them. At moderate field intensities, however, multiple excitation of the clusters dominates over its direct ionization \cite{Laarmann,Kuleff10,DemekhinICD11}. These multiply-excited states  then undergo an  energy transfer  in a very efficient way  to produce differently charged ions.   Such ionization mechanisms  represent a particular case of the broader class of phenomena, known as interatomic Coulombic decay (ICD) \cite{Cederbaum97prlicd}, important in various fields ranging from physics   to biochemistry \cite{Averbukh,Hergenhahn}.

Obviously, the blockade mechanism will strongly influence the efficiency of simultaneous excitation of several identical atoms in a cluster. Here, we investigate how to efficiently  produce multiply-excited homoatomic clusters by a single intense laser pulse, and how the energy of several photons  deposited on different species of a cluster  is then transferred to its  rest. In order to illustrate our findings, we concentrate on a show-case example of the double-excitation of  Ne$_2$ by two photons. On the one hand, Ne$_2$ is amenable to accurate quantum calculations, including all processes evoked by the strong pulse, as well as the  underlying nuclear dynamics, and on the other hand, this example is of   interest to experimentalists by itself. Importantly, it allows for a transparent interpretation of the underlying physics. The present  results allow us to draw  general conclusions which are  valid not only for dimers but also  for larger clusters.

\begin{figure}
\includegraphics[scale=0.40]{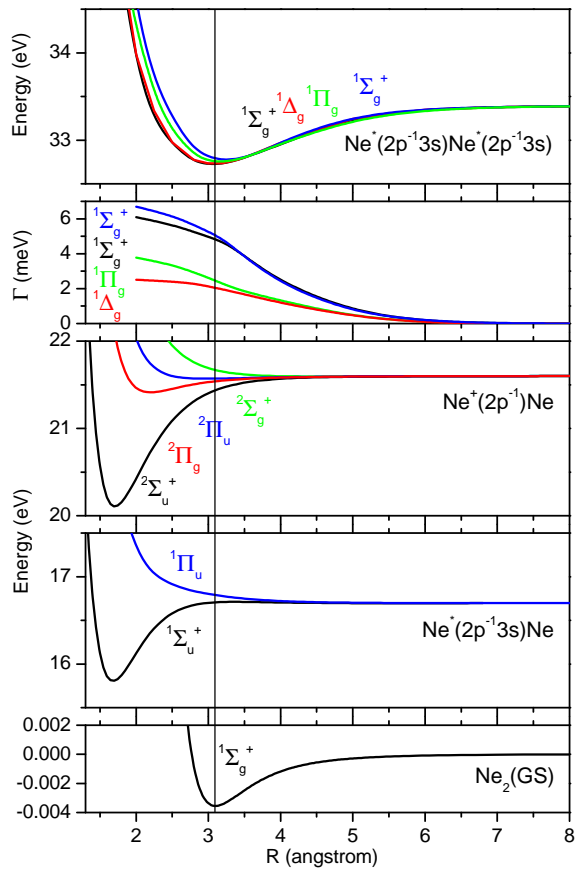}
\caption{(Color online) Presently computed \emph{ab initio} PECs and ICD  rates of the states relevant to the process (\ref{eq:scheme}). These  are the Ne$_2$ ground state, the ungerade Ne$^\ast(2p^{-1}3s)$Ne singly-excited states, the gerade Ne$^\ast(2p^{-1}3s)$Ne$^\ast(2p^{-1}3s)$ doubly-excited states, and the final ionic states Ne$^+(2p^{-1})$Ne. Total decay rates for each doubly-excited state into all final ionic states are shown in the second panel from the top.
}\label{fig_Nedimer}
\end{figure}

For the example chosen the process can be schematically described as follows:
\begin{equation}
\label{eq:scheme}
\mbox{Ne}_2~\stackrel{2\omega}{\longrightarrow}~\left[\mbox{Ne}^\ast(2p^{-1}3s) \right]_2~\stackrel{\mathrm{{ICD}}}{\longrightarrow}~\mbox{Ne}^+(2p^{-1})\mbox{Ne} + e^{-}.
\end{equation}
Two photons of energy $\omega$ excite two Ne atoms in the dimer into their first excited states. These doubly-excited states are located above the first ionization threshold of the dimer  and, thus,  undergo ICD bringing one Ne to its ground state and ionizing the other one. Fig.~\ref{fig_Nedimer} depicts \emph{ab initio}  potential energy curves (PECs) and total ICD  rates for the states relevant for the process (\ref{eq:scheme}) which were computed   as described in detail in Refs.~\cite{trofimov1995,Averbukh05,Kopelke11}. At the equilibrium internuclear distance ($R_e=3.1$~\AA) of the Ne dimer,   resonant population of the  singly-excited states requires an energy of about 16.75~eV. The energy difference between  the statistically weighted average of the two groups of the singly- and the doubly-excited states computed  at 3.1~\AA\, is about 16.03~eV.

How to overcome the enormous blockade of about 0.7~eV and populate the doubly-excited states of the dimer in an efficient way? One can, of course, use two laser pulses with different carrier frequencies of 16.75  and 16.03~eV. Another possibility is to use very short laser pulse of femtosecond or even sub-femtosecond duration, in order to cover the required energy interval. Alternatively, a chirped laser pulse with  carrier frequency changing from  16.75  to 16.03~eV can be applied. These strategies are, however, not feasible at present experimental facilities. We suggest an alternative  strategy of utilizing a single intense pulse to overcome this blockade.

For this purpose we have performed full quantum mechanical calculations on the double-excitation of the neon dimer by intense laser pulses. In order to compute the process~(\ref{eq:scheme}) we combined the previously developed theoretical and computational approaches \cite{Demekhin11SFatom,MolRaSfPRL,DemekhinCO,DemekhinICD11} to evaluate the excitation and decay processes in intense laser fields. The ensuing  non-adiabatic nuclear dynamics has been calculated employing the efficient multiconfiguration time-dependent Hartree (MCTDH) method and code \cite{Meyer90mctdh,MCTDH}.  The values of the electron transition matrix elements for the $2p \to 3s$ excitation, the  $3s \to \varepsilon p$ ionization of the excited states, and the direct two-photon $2p \to  \varepsilon \ell$ ionization of Ne atom were taken from the Refs.~\cite{EL1,EL2,EL3}. Calculations were performed for linearly polarized  laser pulses $\mathcal{E}(t)=\mathcal{E}_0 \,g(t) \cos\omega t$ with Gaussian-shaped   envelopes  $g(t)=e^{-t^2/\tau^2}$ of different   durations $\tau$, carrier frequencies $\omega$, and peak intensities $I_0=\mathcal{E}^2_0/8\pi\alpha$.

\begin{figure}
\includegraphics[scale=0.30]{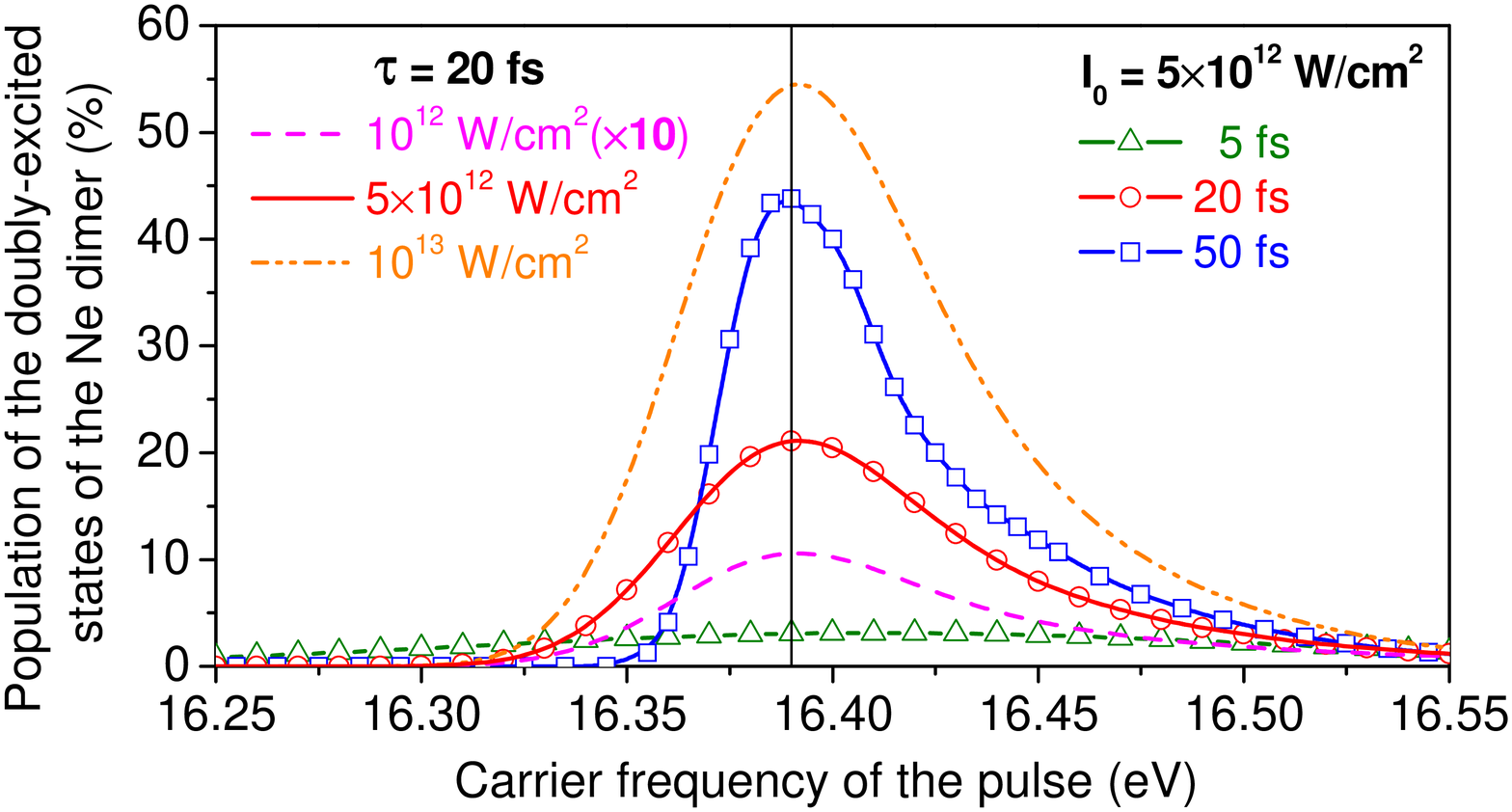}
\caption{(Color online) Final population of the doubly-excited states Ne$^\ast(2p^{-1}3s)$Ne$^\ast(2p^{-1}3s)$ as function  of the carrier frequency $\omega$ after the  Gaussian pulses of different durations $\tau$ and peak intensities $I_0$ have expired. The decay of the doubly-excited states (ICD) has been excluded from the calculations. Open symbols: fixed peak intensity of $5\times10^{12}$~W/cm$^2$ and different pulse durations (see right legend). Curves: fixed pulse duration of 20~fs and different peak intensities (see left legend). Note that some results are shown on an enhanced scale as indicated by the $\times10$ factor in the legend.
}\label{fig_optim}
\end{figure}

Fig.~\ref{fig_optim} depicts  the population of the   doubly-excited states Ne$^\ast(2p^{-1}3s)$Ne$^\ast(2p^{-1}3s)$  as function  of the carrier frequency $\omega$ after expiration of   Gaussian pulses of different durations $\tau$ and peak intensities $I_0$.  In order to preserve the population in the  doubly-excited states, the ICD transitions from them have been excluded from the  calculations. One can see that irrespective of the pulse duration and the peak intensity the optimal carrier frequency of the pulse which allows for a maximal population of the doubly-excited states is equal to $\omega=16.39$~eV. This is exactly the photon energy which is required to access the doubly-excited states by two photons $E_{2exc}=2\omega=32.78$~eV. Another important detail seen from Fig.~\ref{fig_optim} is that the optimum carrier frequency  allows for a very efficient population of the target states (a few tens of percent) at rather moderate peak intensities of about   $10^{13}$~W/cm$^2$, although  it is    detuned far from the resonant transition energies for the first (16.75~eV)   and for the second (16.03~eV)  excitation steps.

\begin{figure}
\includegraphics[scale=0.40]{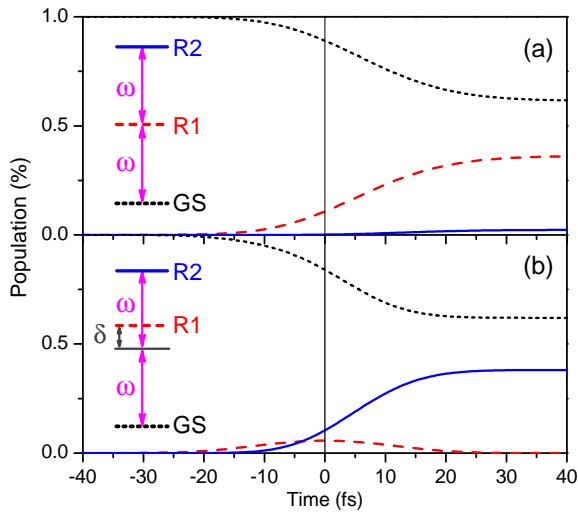}
\caption{(Color online) Results of calculations on a three-level model system exposed to a Gaussian pulse of $\tau=20$~fs duration and carrier frequency $\omega=16.39$~eV. Shown are the populations of the ground state GS (dotted curve), singly-excited state R1 (dashed curve), and doubly-excited state R2 (solid curve) as functions of time. All parameters are taken from the realistic case of  Ne$_2$  considered here: $E_{GS}=0$,  $E_{R2}=2\omega=32.78$~eV, the dipole transition matrix element for the R1$\to$R2 excitation is $D=0.36$~a.u., while the GS$\to$R1 excitation is taken to be twice more probable. Panel (a): the state R1 is resonant for the first excitation step, $E_{R1}=\omega=16.39$~eV, and the system is predominantly singly-excited after the pulse has expired. Panel (b):  the  state R1, $E_{R1}=16.75$~eV, is off-resonant for the first and for the second excitation steps  (detunings are $\delta=\pm 0.36$~eV). Therefore,   R1 serves as a virtual state. The system is essentially only doubly-excited after the pulse has expired. The peak intensities of the pulses are rather moderate and taken to lead to the same level of saturation in both panels (the remaining population of the GS is about 62\%) Of course, one needs larger peak intensity ($7\times10^{12}$~W/cm$^2$) in the non-resonant case (b) compared to the resonant ($1.15\times10^{11}$~W/cm$^2$) case (a).
}\label{fig_model}
\end{figure}

This findings can be rationalized by a simple three-level  model. Let us consider a system with a ground electronic level $\vert GS \rangle$ of energy 0, an intermediate state $\vert R1 \rangle$ of energy $E_{R1}$, and a target state $\vert R2 \rangle$ of energy $E_{R2}$, which is populated by two photons of energy $\omega=E_{R2}/2$.  Fig.~\ref{fig_model} depicts the populations of these levels by a Gaussian pulse of $\tau=20$~fs as a function of time.  The electronic properties of the model  are indicated in the figure  caption and are related to the presently studied Ne$_2$. In    Fig.~\ref{fig_model}(a), the state $\vert R1 \rangle$   is resonant for the first excitation step $E_{R1}=\omega$. The intense pulse manages to transfer  a large fraction of the population from the ground   to the first excited state. During the pulse a small fraction of the singly-excited states is further promoted to the final state by the absorption of the second photon of  resonant energy. After the pulse has expired  the system is predominantly   singly-excited.

\begin{figure}
\includegraphics[scale=0.30]{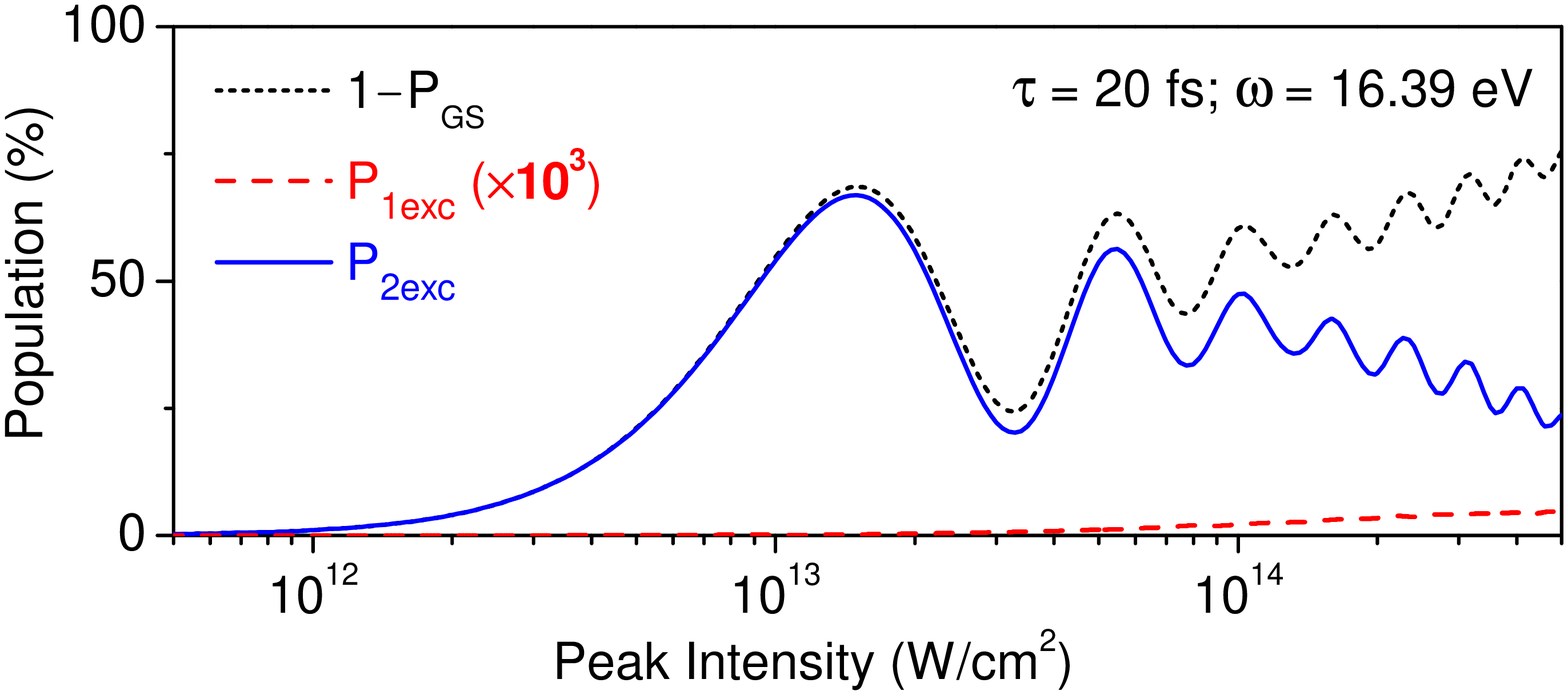}
\caption{(Color online) Final populations of the singly- (dashed curve) and doubly-excited (solid curve) sates of Ne$_2$ after exposure to a Gaussian pulse of 20~fs duration and carrier frequency $\omega=16.39$~eV as functions of the peak intensity of the pulse $I_0$. The decay of the doubly-excited states (ICD) has been excluded from the calculations, but direct ionizations from the singly- and doubly-excited states which take  place during the pulse are accounted for. The dotted curve shows the percentage of the dimers taken out  from their ground state by the pulse. Note that the population of the singly-excited states is shown on an enhanced scale as indicated by the $\times10^3$ factor in the legend and, thus, is negligible at the considered peak intensities. The difference between solid and dotted curves indicates the impact of the direct ionizations which take away   populations from the singly- and doubly-excited states during the pulse.}\label{fig_intens}
\end{figure}

In Fig.~\ref{fig_model}(b), both the first and the second excitation steps are off-resonant, but the whole two-photon transition is resonant.   In this case the intermediate state $E_{R1}$ is populated only virtually, and  its virtual population is  promoted further to the target state $E_{R2}$ during the pulse duration. Very importantly, after the pulse has expired the final population of the virtual intermediate state is negligibly small, and  the system is essentially only doubly-excited. This is due to the energy conservation law which holds for the whole transition $E_{R2}=2\omega$, but does not hold for the first step  $E_{R1}\ne \omega$. Because of the nuclear dynamics and the large number of electronic states participating, the presently studied Ne$_2$ case is much more complicated, but the results shown in Fig.~\ref{fig_optim} confirm the conclusion drawn. In analogy, one may expect for larger clusters that the target states of several excited atoms are optimally populated if the   total energy of several photons fits to the energy of these target states.

Let us now turn back to the  process~(\ref{eq:scheme}). Fig.~\ref{fig_intens} depicts the final populations of the singly- (dashed curve) and doubly-excited (solid curve) sates of Ne$_2$  as functions of the peak intensity of the pulse   after   a Gaussian pulse with  optimal carrier frequency $\omega=16.39$~eV and 20~fs duration has expired. The decay of the doubly-excited states (ICD) has been excluded from the calculations, but direct ionizations from the singly- and doubly-excited states which take  place during the pulse are accounted for. As was already discussed above, the population of the singly-excited states is almost negligible (note the $\times10^3$ factor), and at  peak  intensities below  $10^{13}$~W/cm$^2$ all the dimers which were  taken out  from their ground state by the pulse (dotted curve) are promoted to the doubly-excited states. Interestingly, these dimers constitute  a big fraction (69\% at  $1.5\times 10^{13}$~W/cm$^2$).

\begin{figure}
\includegraphics[scale=0.40]{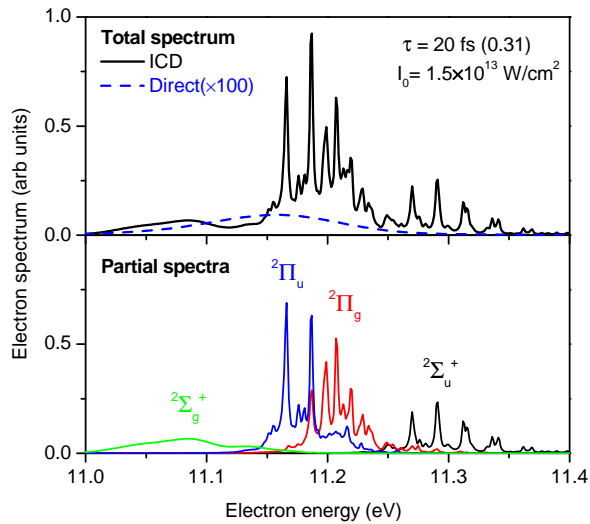}
\caption{(Color online) Computed electron spectra after exposure of Ne$_2$ to a Gaussian pulse of 20~fs duration, carrier frequency $\omega=16.39$~eV, and peak intensity $1.5\times 10^{13}$~W/cm$^2$. Upper panel: Total spectrum. This spectrum is identical with the ICD spectrum because the individual contribution to the spectrum from direct ionization  (broken curve)  is negligible. After the pulse has expired 31\% of the dimers remain in their ground state and 69\% are promoted to the doubly-excited states and undergo ICD (see also Fig.~\ref{fig_intens}). Lower panel: The breakup of the  spectrum into the contributions of the four  final ionic  states (see Fig.~\ref{fig_Nedimer}).}\label{fig_spec}
\end{figure}

At  peak intensities above $10^{13}$~W/cm$^2$   ionization of the dimer starts to play a noticeable role. This is the ionization of the singly-excited states Ne$^\ast(2p^{-1}3s)$Ne by a second photon, and also the ionization of the doubly-excited states Ne$^\ast(2p^{-1}3s)$Ne$^\ast(2p^{-1}3s)$ by a third photon. These `parasitic' processes become important   at    intensities  of about $10^{14}$~W/cm$^2$, and   considerably reduce the `desired' population of the doubly-excited states  (see the difference between solid and dotted curves in  Fig.~\ref{fig_intens}). We also note that the direct two-photon ionization of the ground state is negligible at the considered peak intensities. The maximal population of the target doubly-excited states   by the 20~fs pulse requires   the optimal carrier frequency $\omega=16.39$~eV and   peak intensity of  $I_0=1.5\times 10^{13}$~W/cm$^2$.

Let us now turn to the decay mechanism of the target states and to resulting electron spectra. The average decay lifetime of the Ne$^\ast(2p^{-1}3s)$Ne$^\ast(2p^{-1}3s)$ states at the equilibrium internuclear distance is around 200~fs  (see  total decay rates in   Fig.~\ref{fig_Nedimer}). Consequently, ICD  takes place after the 20 fs pulse has already expired (i.e., in a field-free regime). The total electron spectrum   of Ne$_2$ exposed to the  $\tau=20$~fs pulse of   carrier frequency   $\omega=16.39$~eV and   peak intensity    $I_0=1.5\times 10^{13}$~W/cm$^2$ is depicted in the upper panel of Fig.~\ref{fig_spec} by a solid curve. At this field intensity, the ionization of the dimer proceeds entirely  via the ICD mechanism, since the individual contribution to the spectrum from the direct ionization  (dashed curve shown on an enhanced scale) is negligible. After the pulse has expired, 31\% of the dimers remain in their ground state and the rest is promoted to the decaying states. Therefore, the integral intensity of the spectrum amounts to 69\%.

Interestingly, the computed total electron spectrum in Fig.~\ref{fig_spec} exhibits   well pronounced  resonant structures, which are due to the vibrational levels of the final ionic states Ne$^+(2p^{-1})$Ne. One can learn about the individual ionization pathways  from the lower panel of Fig.~\ref{fig_spec} where the breakup of the spectrum into the individual contributions of the four final ionic states of different symmetry  is depicted. We also mention that these final ionic states are in part weakly bound or even dissociative (see Fig.~\ref{fig_Nedimer}). Following  Fig.~\ref{fig_Nedimer}, the high-energy electrons produced via the decays into the $^2\Sigma^+_u$, $^2\Pi_g$, and partly into the $^2\Pi_u$  final ionic states result in the formation of bound singly-charged  Ne$^+_2$  dimers. The low-energy electrons associated with the decay into the $^2\Sigma^+_g$ and partly into the $^2\Pi_u$    ionic states result in the fragmentation of the dimer into a Ne$^+$ ion and a Ne atom. For larger clusters, several excited pairs are likely to be present and undergo ICD, and hence  fragmentation  or Coulomb explosion  will rather be  the
rule. Here, it also helps that the efficiency of the interatomic decay grows with the size of a cluster \cite{ICDcluster,ICDbulksurf}.

In conclusion, it is demonstrated that the   blockade of  double-excitation of  identical  atoms in a  cluster by the absorption of two photons from the same pulse can be overcome efficiently even for rather long pulses and moderate peak intensities. High-level \emph{ab initio} calculations of the double-excitation of   Ne dimers demonstrate  that the  total energy of the two photons must fit to the energy of the target states, i.e. that the whole double-excitation process is resonant. We suggest that this rule also holds  for the  multi-photon multiple-excitation  of larger clusters. This excitation scheme allows to  prepare   a large fraction of clusters in  multiply-excited states, which then relax  via interatomic decay.  For Ne$_2$ we report \emph{ab initio} electron decay spectra which can directly be compared with experiment. We also note that in molecular clusters the character of the blockade mechanism and the efficient way to overcome it will be   modified owing to the  internal degrees of freedom and  manifold of participating electronic states. Moreover, the subsequent decay process in molecular clusters is, as a rule, much faster \cite{Averbukh,JahnkeNATURAwater}.  Investigation of how the energy of several photons is deposited on different subunits in a system and understanding of how this energy is then transferred to the rest of the system is an issue of general importance in the science of intense radiation, and the present study is a first step in this direction.

\begin{acknowledgements}
The research leading to these results has received funding from the ERC under the EU's  FP7 (AIG No. 227597) and  from the IMPRS at the MPIK, Heidelberg.
\end{acknowledgements}


\begin{thebibliography}{38}
\expandafter\ifx\csname natexlab\endcsname\relax\def\natexlab#1{#1}\fi
\expandafter\ifx\csname bibnamefont\endcsname\relax
  \def\bibnamefont#1{#1}\fi
\expandafter\ifx\csname bibfnamefont\endcsname\relax
  \def\bibfnamefont#1{#1}\fi
\expandafter\ifx\csname citenamefont\endcsname\relax
  \def\citenamefont#1{#1}\fi
\expandafter\ifx\csname url\endcsname\relax
  \def\url#1{\texttt{#1}}\fi
\expandafter\ifx\csname urlprefix\endcsname\relax\def\urlprefix{URL }\fi
\providecommand{\bibinfo}[2]{#2}
\providecommand{\eprint}[2][]{\url{#2}}


\bibitem{Urban_NatPhys09}
E. Urban, \emph{et al.}, Nature Phys. \textbf{5}, 110 (2009).

\bibitem{junct1}
I.\,Giaever\,and\,H.R.\,Zeller,\,Phys.\,Rev.\,Lett.\,\textbf{20},\,1504\,(1968).

\bibitem{junct2}
H.R. Zeller and I. Giaever, Phys. Rev. \textbf{181}, 789 (1969).

\bibitem{junct3}
T.A.\,Fulton\,and\,G.J.\,Dolan,\,Phys.\,Rev.\,Lett.\,\textbf{59},\,109\,(1987).

\bibitem{junct4}
P.J.M.\,van\,Bentum,\,\emph{et\,al.},\,Phys.\,Rev.\,Lett.\,\textbf{60},\,369\,(1988).

\bibitem{junct5}
M.A. Kastner, Rev. Mod. Phys. \textbf{64}, 849 (1992).

\bibitem{BLREV10}
D. Comparat and P. Pillet, J. Opt. Soc. Am. B \textbf{27}, A208 (2010).

\bibitem{primer1}
J.M. Raimond, \emph{et al.}, J. Phys. B \textbf{14}, L655 (1981).

\bibitem{primer2}
W.R. Anderson, \emph{et al.}, Phys. Rev. Lett. \textbf{80}, 249 (1998).

\bibitem{primer3}
I. Mourachko, \emph{et al.}, Phys. Rev. Lett. \textbf{80}, 253 (1998)

\bibitem{primer4}
A. Reinhard, \emph{et al.} Phys. Rev. Lett. \textbf{100}, 123007 (2008).

\bibitem{primer5}
T. Cubel Liebisch, \emph{et al.}, Phys. Rev. Lett. \textbf{95}, 253002 (2005).

\bibitem{primer6}
A. Reinhard, \emph{et al.}, Phys. Rev. A \textbf{78}, 060702(R) (2008).

\bibitem{Zoller_PRL00}
D. Jaksch,  \emph{et al.}, Phys. Rev. Lett. \textbf{85}, 2208 (2000).

\bibitem{Zoller_PRL01}
M. D. Lukin, \emph{et al.}, Phys. Rev. Lett. \textbf{87}, 037901 (2001).

\bibitem{PhotonBL05}
K. M. Birnbaum, \emph{et al.}, Nature \textbf{436}, 87 (2005).


\bibitem{Gaetan_NatPhys09}
A. Ga\"etan,  \emph{et al.}, Nature Phys. \textbf{5}, 115 (2009).


\bibitem{clust1}
O. F. Hagena, Surf. Sci. \textbf{106}, 101 (1981).

\bibitem{clust2}
H. Thomas, \emph{ et al.}, Phys. Rev. Lett. \textbf{108}, 133401 (2012).

\bibitem{FLASH}
W. Ackermann, \emph{et al.},  {Nature photonics} \textbf{1}, 336  (2007).

\bibitem{FERMI}
Home page of  FERMI at Elettra in Trieste, Italy, http://www.elettra.trieste.it/FERMI/.

\bibitem{highharm1}
G. Sansone, \emph{et al.},  {Science} \textbf{314}, 443 (2006).

\bibitem{highharm2}
E. Goulielmakis,  \emph{et. al.},  {Science} \textbf{320}, 1614 (2008).

\bibitem{Saalmann06}
U. Saalmann, \emph{et al.}, J. Phys. B \textbf{39}, R39 (2006).

\bibitem{Laarmann}
T. Laarmann, \emph{et al.}, Phys. Rev. Lett. \textbf{92}, 143401 (2004).

\bibitem{Kuleff10}
A.I. Kuleff, \emph{et al.}, Phys. Rev. Lett. \textbf{105}, 043004 (2010).

\bibitem{DemekhinICD11}
Ph.V. Demekhin, \emph{et al.}, Phys. Rev. Lett. \textbf{107}, 273002 (2011).

\bibitem{Cederbaum97prlicd}
L.S. Cederbaum, \emph{et al.}, Phys. Rev. Lett. \textbf{79},   4778 (1997).

\bibitem{Averbukh}
V. Averbukh, \emph{et al.}, J. Electr. Spectr. Relat. Phen.  \textbf{183}, 36 (2011).

\bibitem{Hergenhahn}
U. Hergenhahn, J. Electr. Spectr. Relat. Phen.  \textbf{184}, 78 (2011);   Int. J. Radiat. Biology (2012) in press DOI:10.3109/09553002.2012.698031.

\bibitem{trofimov1995}
A.B.\,Trofimov\,and\,J.\,Schirmer,\,J.\,Phys.\,B\,\textbf{28},\,2299\,(1995).

\bibitem{Averbukh05}
V. Averbukh and L.S. Cederbaum, J. Chem. Phys. \textbf{123}, 204107 (2005).

\bibitem{Kopelke11}
S. Kopelke, \emph{et al.}, J. Chem. Phys. \textbf{134}, 094107 (2011).

\bibitem{Demekhin11SFatom}
Ph.V. Demekhin and L.S. Cederbaum, Phys. Rev. A  \textbf{83},  023422  (2011).

\bibitem{MolRaSfPRL}
L.S.\,Cederbaum,\,\emph{et\,al.},\,Phys.\,Rev.\,Lett.\,\textbf{106},\,123001\,(2011).

\bibitem{DemekhinCO}
Ph.V. Demekhin, Y.-C. Chiang, and L.S. Cederbaum, Phys. Rev. A \textbf{84} (2011) 033417.

\bibitem{Meyer90mctdh}
H.-D.Meyer, \emph{et al.},  Chem. Phys. Lett. \textbf{165}, 73  (1990).

\bibitem{MCTDH}
G.A. Worth, M.H. Beck,  A. J\"{a}ckle, and H.-D. Meyer. The MCTDH Package, see http://mctdh.uni-hd.de.

\bibitem{EL1}
W.F. Chan, \emph{et al.}, Phys. Rev. A \textbf{45}, 1420 (1992).

\bibitem{EL2}
R. Kau,  \emph{et al.}, J. Phys. B \textbf{29}, 5673 (1996).

\bibitem{EL3}
C. McKenna and H.W. van der Hart, J. Phys. B \textbf{37}, 457 (2004).

\bibitem{ICDcluster}
R. Santra, J. Zobeley, and L.S. Cederbaum, Phys. Rev. B  \textbf{64}, 245104 (2001).

\bibitem{ICDbulksurf}
G. \"{O}hrwall,  \emph{et al.}, Phys. Rev. Lett. \textbf{93}, 173401 (2004).

\bibitem{JahnkeNATURAwater}
T. Jahnke,  \emph{et al.}, Nature Physics \textbf{6}, 139 (2010).

\end{thebibliography}
\end{document}